\documentclass[aps,prb,twocolumn,superscriptaddress,]{revtex4-2}
\usepackage{amsmath}
\usepackage{amssymb}
\usepackage{amsfonts}
\usepackage{graphicx}
\usepackage{xcolor}
\usepackage{hyperref}
\usepackage[caption=false, subrefformat=parens]{subfig}
\usepackage{ifthen}
\usepackage{bbold}
\usepackage{siunitx}

\usepackage{tikz}
\usetikzlibrary{patterns}
\usetikzlibrary{decorations.pathreplacing, calligraphy}
\usepackage{pgfplots}
\DeclareUnicodeCharacter{2212}{−}
\usepgfplotslibrary{groupplots,dateplot}
\usetikzlibrary{patterns,shapes.arrows,backgrounds,intersections}
\pgfplotsset{compat=newest}

\newcommand{\tr}[1][]{\ifthenelse{\equal{#1}{}}{\mathrm{Tr}\,}{\mathrm{Tr}\left[#1\right]}}

\newcommand{\Wg}{\mathrm{Wg}}

\newcommand{\bra}[1]{\langle {#1} |}
\newcommand{\ket}[1]{| {#1} \rangle}
\newcommand{\expect}[1]{\left\langle {#1} \right\rangle}

\begin{document}

%\preprint{}


\title{Disorder Suppression in Topological Semiconductor-Superconductor Junctions}


\newcommand{\cmtc}{Condensed Matter Theory Center and Joint Quantum Institute, Department of Physics, University of Maryland, College Park, Maryland 20742-4111, USA}
\author{Stuart N. Thomas}
\email[]{snthomas@umd.edu}
\affiliation{\cmtc}
%\author{Anton R. Akhmerov}
%\affiliation{Kavli Institute of Nanoscience, Delft University of Technology, P.O. Box 4056, 2600 GA Delft, The Netherlands}
\author{Sankar Das Sarma}
\affiliation{\cmtc}
\author{Jay D. Sau}
\affiliation{\cmtc}

\date{\today}

\begin{abstract}
Disorder in a proximitizing bulk superconductor can scatter quasiparticles in a putative topological superconductor and eventually destroy the topological superconducting state. We use a scattering approach and a random-matrix calculation to estimate the disorder scattering time in a topological Josephson junction. We find that the disorder scattering rate from the bulk of the superconductor, even in the strong coupling limit, is suppressed in the ratio of Fermi momenta between the semiconductor and superconductor. This suppression of disorder scattering is accompanied by near perfect Andreev reflection at such semiconductor/superconductor interfaces, which can be used as a signature of such clean proximity effect. We also find that these results can be understood by a semiclassical estimate of scattering. We discuss limits in other systems such as the semiconductor nanowire where disorder scattering is suppressed according to similar classical estimates.
%Specifically, we calculate an experimentally feasible mean free path for the superconducting metal using physical parameters. In an Al-InAs junction, this value is on the order of $10^{-1}$nm.
\end{abstract}

%\keywords{}

\maketitle

\section{Introduction}%
Majorana zero modes (MZMs) in a topological superconductor provide a possible realization of topological quantum qubits due to their 
non-Abelian braiding statistics~\cite{sarma2015,sau2021topological}.
Recent studies~\cite{lutchyn2010,pientka2017} now predict topological superconductivity in a number of semiconductor-superconductor (SM-SC) heterostructures where the proximity effect from the SC induces a pair potential in the SM.
However, disorder scattering from impurities in both the semiconductor and superconductor can destabilize the topological superconducting phase~\cite{lutchyn2012, potter2011}. In fact, most topological superconductors are destroyed by a disorder back
scattering rate comparable to the superconducting gap. This places a rather stringent constraint on the mean-free path of the SM-SC system 
to exceed the superconducting coherence length~\cite{sau2012, motrunich2001,ahn2021estimating}.

Several theoretical studies~\cite{potter2011,lutchyn2012,hui2014,cole2016,manesco2021mechanisms,kurilovich2022disorder} over the past decade have probed the effect of bulk disorder in the proximity-inducing superconductor on topological phenomena in semiconductors. However, research on the induced topological gap in semiconductors ~\cite{potter2011,lutchyn2012,cole2016,kiendl2019} appears to lead to conflicting results. Analytical calculations~\cite{potter2011,lutchyn2012} show that disorder effects from the proximity-inducing superconductor can only be avoided in the weak tunneling limit between the superconductor and semiconductor. Numerical~\cite{cole2016} and self-consistent Born~\cite{hui2015} calculations suggest these analytical results are truly limited to weak coupling, finding that strong proximity coupling between a disordered superconductor and semiconductor leads to suppression of topological superconductivity. This could be particularly devastating for topological superconductivity in planar Josephson junctions~\cite{pientka2017,ren2019,dartiailh2021phase,banerjee2022}. On the other hand, arguments made in the context of thin-shell superconductors in nanowires~\cite{kiendl2019} suggest that disorder scattering from the proximity-inducing superconductor can be ignored
 in the limit of the Fermi wave-vector mismatch between the superconductor and semiconductor.

In this paper, we use a random-matrix theory approach~\cite{brouwer1996} to understand the density and tunnel coupling dependence of disorder scattering from the proximitizing superconductor. This method will help unify the various limits of superconducting scattering~\cite{potter2011,lutchyn2012,hui2015,cole2016,kiendl2019} that have been studied in the past. The combination of random-matrix theory and semiclassical approaches can overcome the computational challenge associated with the large length-scale difference between the Fermi wavelength of the superconductor and the coherence length in the semiconductor.
%A key advantage over numerical simulations is the ability to accurately portray the Fermi momentum mismatch between the semiconductor and the superconductor. Due to computational limits, this physical ratio is difficult to properly implement in a tight-binding model. Additionally, the scattering formalism removes the need for the self-consistent approximation used in the Born approach.
\begin{figure}[h]
  \subfloat[\label{fig:model} Model of Josephson junction]{
  \begin{tikzpicture}[scale=0.5]
    \draw (0,0) -- (12,0);
    \draw (8,0) -- (8,3);
    \draw (0,3) -- (12,3);
    \draw (4,0) -- (4,3);

    \node[align=center] at (6,4) {$W \ll \xi_{d}$};
    \draw [decorate,
        decoration = {calligraphic brace}, ultra thick] (4,3.2) --  (8,3.2);

    \fill[pattern=dots, pattern color=black!20!white] (8,0) rectangle (12,3);
    \fill[pattern=dots, pattern color=black!20!white] (0,0) rectangle (4,3);
    \node[] at (2,2) {DSC};
    \node[] at (6,2) {SM};
    \node[] at (10,2) {DSC};
    \node[align=center] at (2,1) {$\Delta(x)=\Delta$\\$k_{F}=k_{F,sc}$};
    \node[align=center] at (6,1) {$\Delta(x)=0$\\$k_{F}=k_{F,sm}$};
    \node[align=center] at (10,1) {$\Delta(x)=\Delta$\\$k_{F}=k_{F,sc}$};
  \end{tikzpicture}
  }%
  %\qquad
  \\
  \subfloat[\label{fig:eff-model} Effective model for right interface of JJ.]{
  \begin{tikzpicture}[scale=0.5]
    \draw (0,0) -- (12,0);
    \draw (8,0) -- (8,3);
    \draw (0,3) -- (12,3);
    \draw (4,0) -- (4,3);

    \node[align=center] at (6,4) {$L_{N}=\xi_d$};
    \draw [decorate,
        decoration = {calligraphic brace}, ultra thick] (4,3.2) --  (8,3.2);
    \fill[pattern=dots, pattern color=black!20!white] (4,0) rectangle (8,3);

    \node[] at (2,2) {SM};
    \node[] at (6,2) {N};
    \node[] at (10,2) {SC};
    \node[align=center] at (2,1) {$\Delta(x)=0$\\$k_{F}=k_{F,sm}$};
    \node[align=center] at (6,1) {$\Delta(x)=0$\\$k_{F}=k_{F,sc}$};
    \node[align=center] at (10,1) {$\Delta(x)=\Delta$\\$k_{F}=k_{F,sc}$};
  \end{tikzpicture}
  }
\caption{\label{fig:models} Illustrations of theoretical models. SM represents semiconductor, N represents a diffusive metal, DSC represents a disordered superconductor and SC represents a clean superconductor. $\xi_d$ is the disordered superconducting coherence length, $k_F$ is the Fermi wave-vector and $\Delta$ is the bulk pair potential.}
\end{figure}
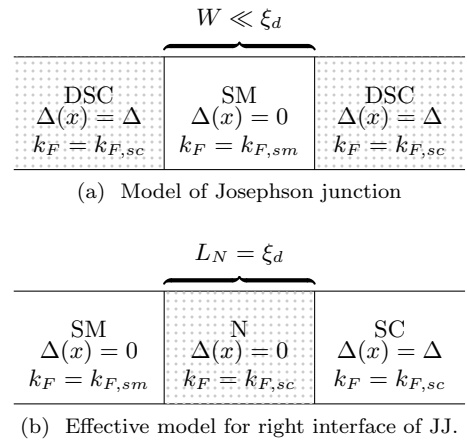

While our work will discuss different platforms of topological superconductivity, we start 
by considering a planar Josephson junction with a short semiconducting region (see Fig.~\ref{fig:model}) since this platform requires strong coupling between the semiconductor and superconductor~\cite{pientka2017}. The strong coupling allows one, in principle, to cross into a topological phase by tuning the phase difference between the superconductors to $\pi$ ~\cite{pientka2017}. Concomitant experimental works show promising evidence for a possible topological superconducting phase in these systems~\cite{ren2019,dartiailh2021phase,fornieri2019,banerjee2022a}.
%The topological superconducting nature of the junction can be understood as a multi-channel topological superconductor where the $p-$wave gap, $\Delta_p$, is larger than the $s-$wave gap $|\Delta_s|<|\Delta_p|$, which is the spectral gap of the clean system. 
An analysis of this junction has shown that disorder enhances the spectral gap near $k\sim k_F$, while suppressing the topological gap near $k\sim 0$~\cite{haim2019}, suggesting a topological phase for a strongly spin-orbit coupled Josephson junction in the limit where the disorder scattering is smaller than the microscopic superconducting gap. This result is consistent with a similar result for topologically superconducting semiconductor nanowires in the strong-spin-orbit coupled regime~\cite{sau2012} and confirms that strong disorder scattering, which could potentially arise from a strongly disordered proximitizing superconductor, would destroy topological superconductivity in both platforms.

The proximity-induced superconducting pairing potential $\Delta$ in the semiconductor in Fig.~\ref{fig:model} is related to the Andreev reflection rate from the superconductor, which can be written as $\Delta\sim\Omega_{sc}|r_{eh}|^2$, where $|r_{eh}|^2$ is the Andreev reflection probability and $\Omega_{sc}$ is the scattering rate for electrons in the semiconductor reflecting between the superconductors.
%We can write the total scattering rate $\Omega_{sc}$ in terms of the delay times associated with, $\tau_{sm}$, traversing the semiconductor and the mean reflection time from the superconductor, yielding $\Omega_{sc}=[\tau_{sm}+\tau_{A}|r_A|^2 + \tau_{N}|r_{ee}|^{2}]^{-1}$ where  $\tau_{N}$ is the average time of normal reflection, and $\tau_{A}\sim\Delta_{s}^{-1}$ \cite{brouwer1997quantum} is the Andreev reflection time.
The presence of disorder in the superconductor introduces scattering at a rate $\tau_{d}^{-1}$ that breaks momentum conservation in the semiconductor. We find that this scattering is dominated by normal scattering with the probability $|r_{ee}|^2$ (see Appendix~\ref{sec:offdiagonal}), which is entirely momentum conservation breaking in the case of a transparent semiconductor-superconductor interface. We can then approximate the scattering rate $\tau_{d}^{-1}$ as the rate of normal scattering $\tau_d^{-1}\sim\Omega_{sc}|r_{ee}|^2$, written in terms of the induced pairing potential as
%is therefore a product of the total scattering rate $\Omega_{sc}$ in the Josephson junction with the probability $|r_{ee}|^{2}$ of normal reflection. Taking the first order in $|r_{ee}|^{2}$, the disorder scattering rate becomes $\tau_{d}^{-1}\sim |r_{ee}|^{2}\Delta_{s}/(\tau_{sm}\Delta_{s}+1)$. Since $\Delta_{s}\sim\Delta_{0}\sim\Delta$,
% \begin{equation}
%   \label{eq:tau_d}
%   \tau_{d}^{-1}\sim \Omega_{sc}|r_{ee}|^2\approx \Omega_{sc}|r_{ee}|^2/|r_{eh}|^2\sim\Delta |r_{ee}|^{2}.
% \end{equation}
% \begin{equation}
%   \label{eq:tau_d}
%   \tau_{d}^{-1}\sim \Delta|r_{ee}|^{2}.
% \end{equation}
\begin{equation}
  \label{eq:tau_d}
  % \Delta \tau_{d}\sim \frac{1-|r_{ee}|^{2}}{|r_{ee}|^{2}}.
\tau_{d}^{-1}\sim \frac{ \Delta |r_{ee}|^{2}}{1-|r_{ee}|^{2}}.
\end{equation}

The disorder scattering in Eq.~\ref{eq:tau_d}, with the condition of a large topological gap, i.e. $\tau_d\Delta \gg 1$~\cite{sau2012}, leads to the requirement $|r_{ee}|^{2}\ll 1$. This condition can be experimentally verified by checking for near perfect Andreev reflection, i.e. $G_{NS}\sim  (4e^2/h) N_{sm}$. However, the possibility of weak normal reflection from the superconductor in the strong coupling regime is unclear due to the apparent contradiction between the numerical results of Cole et al~\cite{cole2016} and the Fermi momentum mismatch arguments in Kiendl et al~\cite{kiendl2019}.
In our work, we show that disorder scattering  can indeed be suppressed at such interfaces with a strong Fermi level mismatch.
%as well as the correct conditions of width of the semiconductor. 
We also estimate disorder scattering in other systems such as a thin superconductor \cite{kiendl2019} and a finite-length semiconductor \cite{lutchyn2012, potter2011}, applying a semiclassical estimate of scattering to these situations. We use these arguments to provide a unifying view of previous works on the subject~\cite{potter2011,lutchyn2012,cole2016,kiendl2019}. Finally, we discuss how the disorder scattering maybe restored in the presence of a magnetic field and how our results compare to numerical simulation.

%OUTLINE
The paper is organized as follows.
In Sec. ~\ref{sec:full-calculation}, we introduce an effective model and use a random matrix theory approach to solve for the scattering probabilities.
In Sec.~\ref{sec:physical-realization}, we insert physical parameters to achieve a lower bound for the mean free path.
In Sec.~\ref{sec:classical-estimate}, we study additional physical systems and compare with previous literature, applying a semiclassical approximation which is motivated by the scattering result.
In Sec.~\ref{sec:magnetic-field}, we explore the breaking of disorder suppression using a magnetic field and in Sec.~\ref{sec:kwant} we support this notion with numerical results.
Finally, we give a conclusion in Sec.~\ref{sec:discussion}.

\section{Random-matrix theory approach to scattering rate}
\label{sec:full-calculation}
To model the disordered superconductor, we make a simplifying assumption by separating this region into a diffusive normal region of length $L_N$ coupled to a clean superconductor (as shown in Fig.~\ref{fig:eff-model}). The length $L_{N}$ is chosen as $L_N=\xi_{d}$ since a shorter diffusive region (i.e. $L_N\ll \xi_{d}$) underestimates the effect of disorder while a longer region (i.e. $L_N\gg\xi_{d}$) exhibits spurious subgap states~\cite{belzig2001}. Such an approximation to the disordered superconductor retains its essential disorder scattering as well as Andreev reflection processes while preserving the gap.
% This work focuses on an interface between a semiconductor with low electron density and a diffusive metal/superconductor with high density.
The near-perfect Andreev reflection, even in the absence of disorder, is possible only if the superconductor is connected to the semiconductor by an adiabatic (i.e. non-reflective) potential.
We model this adiabatic potential step as a scattering matrix with reflection and transmission amplitudes of either $0$ or $1$ and the transverse wave-functions as eigenvectors.
These assumptions lead to the effective model shown in Fig.~\ref{fig:eff-model}.

To solve for the normal reflection probability $|r_{ee}|^{2}$, and therefore the effective scattering rate according to Eq.~\ref{eq:tau_d}, we compose the scattering matrices of the SM-N interface, the diffusive region and the superconductor.
We first consider the $N_{sm}\times N_{sm}$ reflection matrix $r_L$ of the normal part of the system (i.e. without the superconducting lead) from the the semiconductor side. This matrix depends on $r_{N}$, the $N \times N$ reflection matrix of the diffusive metal. Due to time-reversal invariance, $r_{N}$ is symmetric and therefore admits a polar decomposition
\begin{equation}
\label{eq:polar_decomposition}
    r_N = U^T \sqrt{R_{N}} U,
\end{equation}
where $R_{N}$ is a diagonal matrix of reflection probabilities and $U$ is a unitary matrix. We also define the reflection matrix $r_{sm}$, describing the adiabatic scattering at the  semiconductor-metal interface, as
\begin{equation}
    \label{eq:adiabatic_sm}
    r_{sm} = \sum_i^{N} \Theta \left(k_{i}^2-k_{F,sm}^2\right) \ket{i}\bra{i}.
\end{equation}
Here, $k_{i}$ is the transverse momentum in the $i$-th mode of the metal, $k_{F,sm}$ is the Fermi momentum in the semiconductor and $\Theta$ is the Heaviside step function. Using these two reflection matrices, we can calculate $r_{L}$ as \footnote{A reduction to the nonzero $N_{sm}\times N_{sm}$ block is implied.}
\begin{equation}
\label{eq:rl}
  r_{L}= t_{sm}\left(1-r_{N}r_{sm}\right)^{-1}r_{N}t_{sm}.
\end{equation}
where $t_{sm}\equiv 1-r_{sm}$.

%Generally, the conductance of an SC heterojunction is $G_{NS} = (4e^2/h) \; \tr r_{eh}^\dag r_{eh}$. Taking the trace of $r_{eh}^\dag r_{eh}$ we calculate the conductance of the SM-N-SC junction as
Following Beenakker~\cite{beenakker1992}, the zero-energy conductance resulting from Andreev reflection of electrons coming from the left lead of Fig.~\ref{fig:eff-model} can be written in terms of the eigenvalues of $r_Lr_L^\dag$, ${R_L}_i$, as:
\begin{align}
\label{eq:beenakker}
 G_{NS} = \frac{4 e^2}{h} \sum_i \left(\frac{1-{R_L}_i}{1+{R_L}_i}\right)^2,
\end{align}
%This equation generalizes Beenakker's equation~\cite{beenakker1992} to unequal numbers of left- and right-side modes.
which is bounded by
\begin{equation}
\label{eq:beenakker_avg}
 G_{NS} > \frac{4 e^2}{h}\left( N_{sm} - 4 \tr[r_L r_L^\dag] \right).
\end{equation}
This bound becomes an accurate estimate in the limit of small $R_L$, which we will find applies in limit of small $N_{sm}/N$.

The trace in this conductance is a self-averaging quantity in the limit of large $N_{sm}$, so we can approximate its value by averaging over disorder realizations. This value is equivalent to averaging the trace by choosing the total scattering matrix $s_N$ from a distribution of random matrices in the COE ensemble.
The equivalent channel assumption, motivated by Dorokhov's model~\cite{dorokhov1988} of disordered conductors, implies $U$ (see Eq.~\ref{eq:polar_decomposition}) is uniformly distributed over the unitary group~\cite{beenakker1997}.
With this formula in mind, we write expectation value of the trace $\tr r_{L} r_{L}^{\dagger}$ as
\begin{equation}
\label{eqrL}
    \left\langle \tr r_{L} r_{L}^{\dagger} \right\rangle = \sum_{n=0}^\infty \Gamma_{n}
\end{equation}
where
\begin{align*}
  \Gamma_{n}= &\left\langle \tr t_{sm}U^{T}\sqrt{R_N}U\left(r_{sm} U^{T}\sqrt{R_N}U\right)^n \right.\\
         &\times \left. t_{sm}  \left(U^{\dag}\sqrt{R_N}U^{*}r_{sm}\right)^n U^{\dag}\sqrt{R_N}U^{*} \right\rangle
\end{align*}
noting that terms with unequal $U$ and $U^{*}$ factors vanish~\cite{collins2006}. Here, $t_{sm} =1-r_{sm}$.

Traces over products of random unitary matrices, similar to the above equation have appeared in studies of conductance of 
disordered metal-superconductor junctions and a chaotic quantum dots as well as other systems~\cite{samuelsson2002,barbosa2005}. Diagrammatic methods~\cite{brouwer1996} can evaluate such averages in the large $N$ limit.
We use a similar method to calculate $\Gamma_{n}$ to leading order in $N$ and $N_{sm}$ (see Appendix~\ref{sec:calculation-details} for details):
\begin{equation}
  \label{eq:gamma-final}
   \Gamma_{n} = \frac{N_{sm}^{2}}{N} {\overline{R_{N}}\,}^{n+1} \left(1-\frac{N_{sm}}{N}\right)^{n}.
\end{equation}
Taking the sum over the geometric series in $n$ (see Eq.~\ref{eqrL}) then yields
\begin{align}
  \label{eq:rmat}
    \tr[r_{L} r_{L}^{\dag}] = N_{sm}\,\frac{Z^{2}}{1+Z^{2}}, && Z^{2} = \frac{N_{sm}}{N} \left(\overline{T}_{N}^{\,-1} -1\right)
\end{align}
where $\overline{T}_{N}= 1 - \overline{R_{N}}$ is the average transmission eigenvalue of the normal region. Physically, $Z$ represents the unitless strength of an equivalent $\delta$-function barrier which completely describes the scattering probability of the SM-N junction. Note that this expression is independent of correlations between transmission eigenvalues.
Substituting this into the bound in Eq.~\ref{eq:beenakker_avg} yields an estimate for the bound on the conductance:
\begin{equation}
  \label{eq:conductance_bound_Z2}
    G_{NS} \gtrsim \frac{4 e^{2}}{h} \, N_{sm}\left(\frac{1-3Z^2}{1+Z^{2}}\right).
\end{equation}

The above conductance, which arises entirely from Andreev reflection, provides an upper bound to the normal reflection probability in the superconducting junction:
\begin{equation}
  \label{eq:result}
  |r_{ee}|^2 < \frac{4Z^2}{1+Z^2}.
\end{equation} 
The total disorder scattering from the bulk superconductor is a combination of the above normal reflection as well as the off-diagonal part of the Andreev reflection matrix. As shown in Appendix~\ref{sec:beenakker_generalization}, the full Andreev reflection matrix for time-reversal invariant systems can be written in terms of $r_L r_L^\dagger$ as:
\begin{align}
\label{eq:reh}
 r_{eh} = -i \left(\frac{1-r_L r_L^\dag}{1+r_L r_L^\dag}\right).
\end{align}
% Because of the channel rotation symmetry of the random matrix measure, the mean of the matrix $R_L=r_L r_L^\dag$ is approximately proportional to the identity matrix, a positive definite matrix,
% which becomes suppressed as $Z^2\rightarrow 0$ (Eq.~\ref{eq:rmat}). Therefore
Since $r_{L}r_{L}^{\dag}$ is a positive definite matrix that is suppressed as $Z^{2}\rightarrow0$, we can approximate $r_{eh}\sim -i(1-2 r_Lr_{L}^{\dag})$.
The off-diagonal components are thus bounded by the fluctuations of $2 r_Lr_L^{\dag}$, i.e. $2\sqrt{\tr[R_L^2]/N_{sm}}$, where $R_{L}$ is the diagonal matrix of $r_{L}r_{L}^{\dag}$ eigenvalues. These fluctuations scale as $2 Z^2/(1+Z^{2})$, similar to the mean in Eq.~\ref{eq:rmat} (see Appendix \ref{sec:offdiagonal}).
%We estimate this part of the Andreev reflection by calculating the trace $\langle \tr r_{L} r_{L}^{\dag}r_{L}r_{L}^{\dag}\rangle$. This, in turn, is equivalent to the variance of the reflection eigenvalues ${R_{L}}_{i}$, which leads to  $|r_{od}|^2\sim Z^4$ for small $Z^{2}$ (see Appendix~\ref{sec:offdiagonal}).
% \begin{equation}
%     \label{eq:offdiagonalscattering}
%     |r_{eh}^{od}|^2\sim \left(\frac{Z^2}{1+Z^2}\right)^2.
% \end{equation}
The off-diagonal Andreev reflection probability then scales as $|r_{od}|^2\sim Z^4$, which is thus negligible compared to the normal reflection probability $|r_{ee}|^{2}\sim Z^2$ , corroborating the assumption made in the expression for $\tau_{d}^{-1}$ (Eq.~\ref{eq:tau_d}). This also gives an estimate of the perfect (diagonal) Andreev reflection by substituting $r_{L}r_{L}^{\dagger}\rightarrow Z^{2}/(1+Z^{2})$:
\begin{equation}
 \label{eq:beenakker_approx_prob}
  |r_{eh}^{p}|^{2} \approx \frac{1}{\left(1+2Z^{2}\right)^{2}}.
\end{equation}

\section{Scattering rates at real material interfaces}
\label{sec:physical-realization}
\newcommand{\D}{\mathcal{D}}
To compare with realistic material properties, we connect the Andreev reflection probability to physical parameters. The disordered superconducting coherence length is $\xi_{d}=\sqrt{\xi_{0} \lambda}$ where $\xi_{0}=v_{F,sc}/\Delta$ is the clean superconducting coherence length and $\lambda$ is the mean free path.
In the diffusive regime for the normal region $N$ in Fig.~\ref{fig:eff-model}, the DMPK equation~\cite{dorokhov1988, mello1988, macedo1992} gives an average transmission eigenvalue of $\overline{T}_{N}=\lambda/L_{N}\ll 1$ in the many-channel limit. Setting the normal region length to the disordered superconducting coherence length $\xi_{d}$ (as shown in Fig.~\ref{fig:eff-model}) yields
\begin{equation}
  \label{eq:diffusive}
  Z^{2} \approx \frac{N_{sm}}{N} \sqrt{\frac{\xi_{0}}{\lambda}}.
  % Z^{2} \approx \frac{N_{sm}\xi}{N\lambda}.
\end{equation}
% Eqs.~\ref{eq:result} and \ref{eq:diffusive} form the main technical result of this letter.
This result is valid under the parameter hierarchy $N\gg N_{sm}\,\sqrt{\xi_{0}/\lambda}\gg N_{sm}\gg 1$.
\begin{figure}
  \centering
  \includegraphics[width=0.43\textwidth]{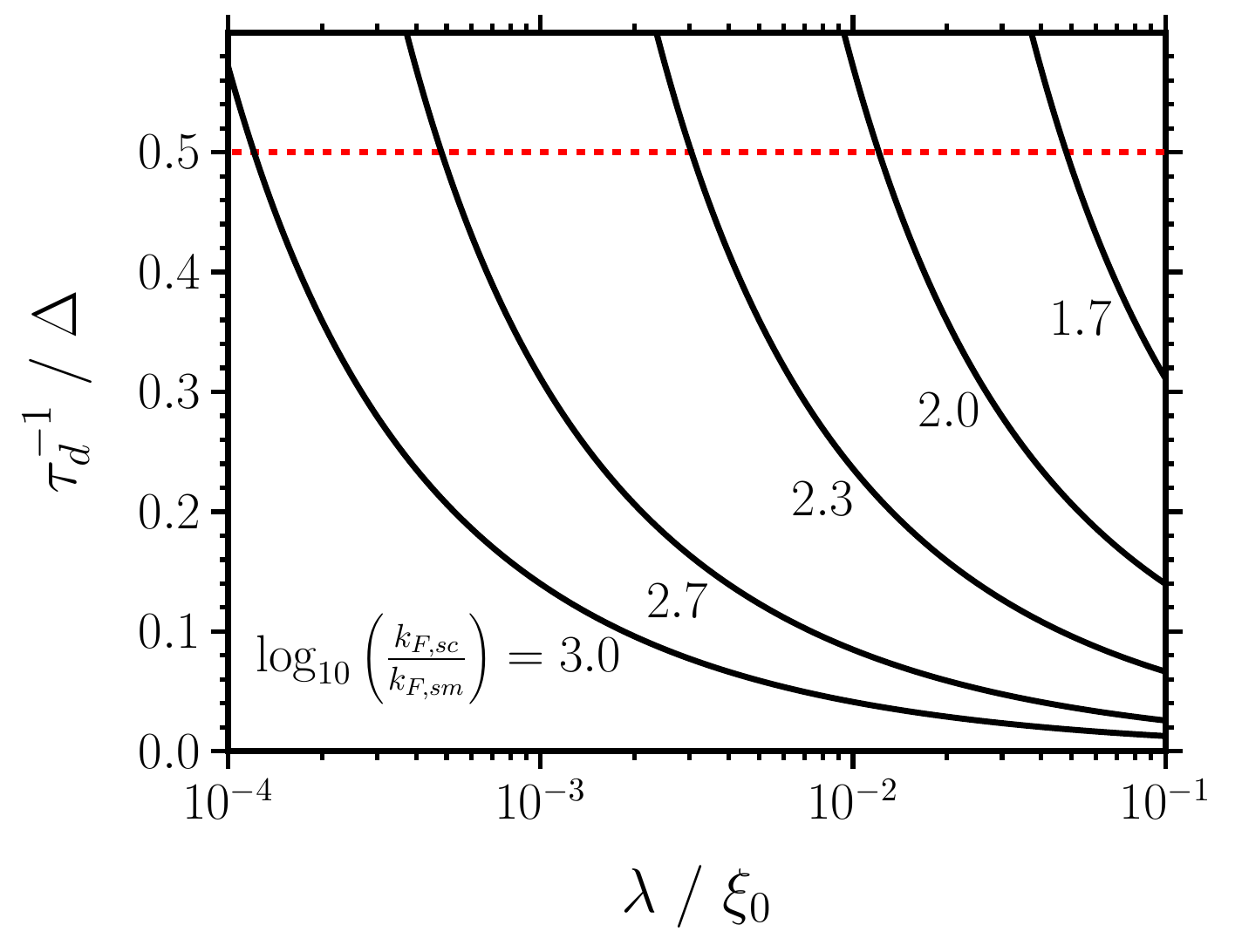}
  \caption{\label{fig:plot} Upper bound of disorder scattering rate in units of the induced gap as a function of mean free path in units of the superconducting coherence length, from Eq.~\ref{eq:result}. Dashed horizontal line represents the maximum rate ($\tau_{d}^{-1}=0.5\Delta$, see~\cite{sau2012}) where a topological gap is possible. The series labeled with $\log_{10}\left(k_{F,sc}/k_{F,sm}\right)=2.7$ and $3.0$ represent InAs-Al and InSb-Al junctions respectively \cite{sladek1957, ashcroft1976}.}
\end{figure}
Fig.~\ref{fig:plot} shows the upper-bounded disorder scattering rate $\tau_d^{-1}$ as a function of mean-free path for various values of the density mismatch at the semiconductor-superconductor interface.
By approximating the ratio $N_{sm}/N \approx k_{F,sm}/k_{F,sc}$ ($k_{F,sm}^{2}/k_{F,sc}^2$ in 3D), we write
\begin{equation}
  \label{eq:physical_Z2}
  % Z^{2} = \frac{k_{F,sm}}{k_{F,sc}} \sqrt{\frac{v_{F,sc}}{3\Delta\lambda}}.
  Z^{2}= \sqrt{ 2\; \frac{m^{*}_{sm}}{m^{*}_{sc}} \; \frac{\mu_{sm}}{\Delta} \; \frac{1}{k_{F,sc}\lambda} }.
\end{equation}
Here, $m^{*}_{sm(sc)}$ is the effective mass in the semiconductor (superconductor), $\mu_{sm}$ is the Fermi energy in the semiconductor and $k_{F,sc}$ is the Fermi momentum in the superconductor.

% Now we consider the length of the junction $L$. To accurately model a disordered superconductor, we restrict the diffusive region to be greater than the disordered superconducting coherence length $\xi$, which is defined as $\sqrt{\mathcal{D}/\Delta}$  We also keep the length shorter than the Thouless length to inhibit corrections to Beenakker's reflection formula~\cite{belzig2001}. Since the Thouless length is also given by $\sqrt{\mathcal{D}/\Delta}$, we set our length $L$ to this value and the ratio $L/\lambda$ becomes $\sqrt{v_{F,sc}/3\Delta \lambda}$ which dominates over the $1$ term:

To produce an induced topological gap in the semiconductor, the disorder scattering time must satisfy $\Delta \tau_{d} \gg 1.08$~\cite{sau2012}. To ensure this inequality is strongly satisfied, we enforce $\Delta \tau_{d}>2$. Combining Eqs.~\ref{eq:tau_d}, \ref{eq:result} and \ref{eq:physical_Z2} gives a resulting minimum bound for the mean free path:
\begin{equation}
  \label{eq:physical_mfp}
  \lambda \gtrsim 242\, \frac{m_{sm}^{*}}{m_{sc}^{*}} \; \frac{\mu_{sm}}{\Delta} k_{F,sc}^{-1}.
\end{equation}
To evaluate the validity of this inequality, we insert specific values for InAs, InSb and Al. For $\mu_{sm}$ we use $46\si{K}$ for InAs and $14\si{K}$ for InSb. Taking the remaining physical values from~\cite{sladek1957} and~\cite{ashcroft1976}, we find that the lower bound on the mean free path is $4.39\si{nm}$ for InAs and $1.07 \si{nm}$ for InSb. The actual requirements on the mean-free path in the superconductor depend on the Fermi energy in the semiconductor $\mu_{sm}$ in a particular device. Planar Josephson junction set-ups~\cite{ren2019} can be topological at higher values of Fermi energy such as the one~\cite{sladek1957} used for the above estimate. Nanowire systems for topological superconductors have Fermi energy that is typically lower than a few K~\cite{lutchyn2010,oreg2010helical}.

Another natural possibility that can arise in some material systems is that the high density superconductor is in the three dimensional limit, while the semiconductor is in the two dimensional limit. In this case the ratio becomes $N_{sm}/N=k_{F,sm}/k_{F,sc}^2 W$, where $W$ is the thickness of the superconductor. This change can further enhance the channel mismatch effect where the $Z^2$ in Eq.~\ref{eq:physical_Z2} is reduced by a factor of $(k_{F,sc}W)^{1/2}$ and enhance the robustness to disorder scattering. Whether this occurs in a planar Josephson junction device depends on the density in the semiconductor that is directly under the superconductor.

\subsection{Application to thin-shell superconductors around nanowires}
\label{sec:thinshell}
Though the calculation in the previous section specifically pertains to planar Josephson junctions with an ideally infinitely long superconductor (Fig.~\ref{fig:model}), we find that the result in Eq.~\ref{eq:result} also applies to a system where the superconductor is a thin-shell around a nanowire~\cite{krogstrup2015,chang2015,kiendl2019}, provided the Fermi wave-vector $k_{F,sc}$ of the superconductor continues to be parametrically large.
The thin-shell superconductor, which is used in many of the experiments on Majorana nanowires~\cite{deng2016,albrecht2016,gul2018ballistic}, differs from the configuration in Fig.~\ref{fig:model} in that the superconductor thickness $d$ is much smaller than the disordered superconducting coherence length. The effective model in Fig.~\ref{fig:eff-model} with a na\"ively shortened diffusive normal region $L_{N}=d$ does not properly exhibit the scattering processes of such a shell. The scattering in the effective model is much weaker than the physical thin-shell system because it does not represent reflection from the surface of the thin-shell.

To represent scattering from the surface, we introduce a barrier of fixed transparency $\Gamma=d/ \xi_{d}$ between the normal metal (N) and the superconductor (S) in Fig.~\ref{fig:eff-model}. This barrier, which represents the reflection from the outside surface of the thin-shell, ensures that the electron propagates a total distance of approximately $\xi_{d}$ before entering the superconducting region, simulating the Brownian path within the superconducting shell.

Since Eqs.~\ref{eq:rl} through \ref{eq:reh} hold for any symmetric reflection matrix $r_{N}$, we can replace the average transmission $\overline{T_{N}}$ in Eq.~\ref{eq:rmat} with the metal-barrier average transmission $\overline{T_{NB}}$ to calculate $|r_{ee}|^{2}$ in the thin-shell system. While the DMPK equation provides a simple approximation for $\overline{T_{N}}$, the calculation of $\overline{T_{NB}}$ requires a composition of scattering matrices, formulated as $\overline{T_{NB}}=\tr t_{NB}^{\dag} t_{NB}/N$ where
\begin{equation}
  \label{eq:thinshell}
  t_{NB} = \sqrt{\Gamma} \left( 1-r_{N}' \sqrt{1-\Gamma}\right)^{-1} t_{N}'.
\end{equation}
Here, $r_{N}'$ and $t_{N}'$ are respectively the right-side reflection matrix and left-to-right transmission matrix of the normal metal region with length $d$. Since $\overline{T_{NB}}$ is self-averaging, we use the polar decomposition and integrate over the unitary group, calculating $\overline{T_{NB}} = (\overline{T_{N}'}^{-1}+\Gamma^{-1} - 1)^{-1}$ to leading order in $N$ where $\overline{T_{N}'}=\tr {t_{N}'}^{\dag}t_{N}'/N$ (see Appendix~\ref{sec:thinshelldetails}). By applying the DMPK equation to $\overline{T_{N}'}$, we find that the transmission $\overline{T_{NB}}$ has a lower bound of $\overline{T_{N}'}\Gamma = \sqrt{\lambda / \xi_{0}}$ which is equivalent to the transmission value found in Eq.~\ref{eq:diffusive}. Therefore the bound on $|r_{ee}|^{2}$ in Eq.~\ref{eq:result} also applies to the superconducting shell system, indicating that perfect Andreev reflection with low disorder scattering can occur in thin superconducting shells where the mean free path satisfies Eq.~\ref{eq:physical_mfp}.

\section{Classical estimate of scattering rates}%
\label{sec:classical-estimate}
Interestingly, the results for both the planar Josephson junction as well as the thin superconducting shell turn out to be qualitatively consistent with a semi-classical estimate of Andreev scattering.  This estimate is similar to the Drude estimate of the conductivity of metals which views electrons as classical particles executing a random walk as a result of collisions with impurities. Andreev reflection is described by electrons becoming holes which retrace the classical path. Ref.~\cite{kiendl2019} finds that this semi-classical argument correctly estimates the proximity-induced pairing from a disordered superconducting shell.

The semi-classical model for scattering in a superconductor is based on the observation that so-called Diffuson and Cooperon diagrams~\cite{evers2008anderson} dominate the long-ranged contributions to propagation of electron waves in a diffusive normal metal. Andreev reflection can be added to this model by connecting the two legs of the Cooperon containing time-reversed electron trajectories by a reflection process~\cite{hekking1994,kurilovich2022disorder}. The resulting picture is similar to the Drude theory of conductance, which assumes that the electron follows a Brownian path that randomly  switches direction after traveling distance $\lambda$ between scattering events, before it Andreev reflects into a hole after traveling a total distance $\xi_{0}$ in the superconductor~\cite{kiendl2019}. The distribution of the electron density following the Brownian motion matches the profile obtained from Diffusion and Cooperon diagrams that obey a diffusion equation. The total penetration of the electron trajectory into the diffusive superconductor is therefore $\xi_{d}=\sqrt{\xi_{0}\lambda}$. Within this classical model, the Andreev reflected hole exactly retraces the time-reversed path of the electron and thus does not contribute to scattering in momentum space. Scattering events where an electron is within a mean-free path $\lambda$ from the interface can lead to an electron reflecting back into the semiconductor.  Thus, each electron suffers $(\xi_{0}/ \lambda) \times (\lambda/ \xi_{d}) = \sqrt{\xi_{0}/ \lambda}$ impurity collisions where it has a chance to return to the semiconductor.
The electron can only return if the momentum after scattering is compatible with the allowed momenta in the semiconductor. Combining these two facts, the total probability of a scattered electron reflecting into the semiconductor is:
\begin{equation*}
 P\left( e\rightarrow e \right) =  \frac{2}{\pi} \,\frac{k_{F,sm}}{k_{F,sc}} \,\sqrt{\frac{\xi_{0}}{\lambda}}.
\end{equation*}
This matches (modulo a numerical pre-factor) the result of the random matrix calculation in Eqs.~\ref{eq:result} and \ref{eq:diffusive} to first order.

We note that these results agree despite the fact that Eq.~\ref{eq:diffusive} is based on the set-up in Fig.~\ref{fig:eff-model}. The key difference between these two systems is that in the effective model, Andreev reflection can happen only in at the boundary of the normal region while in the classical estimate, it can occur anywhere on the diffusive path. This difference has no effect on the reflection probabilities provided that the length of the normal region $L_{N}$ is chosen to be the average penetration length $\xi_{d}$ into the superconductor. Additionally, the momentum non-conserving part of the Andreev reflection, which vanishes in the classical estimate, is sub-leading $Z^{2}$ according to the result in Appendix~\ref{sec:offdiagonal}.

The classical estimate above provides intuition for suppression of SC disorder scattering found in these calculations. In the classical picture, it is quite unlikely for an electron to return to the semiconductor unless it Andreev reflects into a hole and retraces its path backwards in time (an assumption also made in Ref.~\onlinecite{kiendl2019}), suppressing disordered reflection.
Specifically, electrons in the normal metal must have a transverse momentum less than the Fermi momentum of the semiconductor to transmit across the junction (see Eq.~\ref{eq:adiabatic_sm}). If the angle between the particle's momentum and the wire's length is $\theta$, this condition is equivalent to the inequality $k_{F,sc} \sin{\theta} < k_{F,sm}$. When the ratio of Fermi momenta is small, probability of transmission back into the semiconductor becomes $\left( 2/ \pi  \right)k_{F,sm}/k_{F,sc}$ in 2D systems. The consistency between the random matrix result (Eqs.~\ref{eq:result} and \ref{eq:diffusive}) and the classical estimate provide confidence in the scattering rate estimated in this work.

\subsection{Scattering rates in nanowires}
The similarity between the classical estimate and our quantum calculations motivates us to consider other cases such as the proximity-induced superconducting quantum well considered in Refs.~\cite{lutchyn2012, potter2011}. In this system, a barrier at the SM-N interface confines the semiconductor states to quantum well states, which have a broadening $\gamma$ from tunneling into the superconductor in its normal state. This broadening sets the proximity pairing in the semiconductor~\cite{stanescu2010}.
The parameter $\gamma$ is also key to allowing the electrons to scatter back from the diffusive superconductor into the broadened band of the quantum well since it is proportional to the phase space in the semiconductor. Assuming a three-dimensional system, the momentum broadening of states in the semiconductor results in a total phase space proportional to $k_{F,sm}\, \gamma/v_{F,sm}=m^{*}_{sm}\gamma$ for a single band. Comparing this to the phase space area in the superconductor which is proportional to $k_{F,sc}^{2}$ allows us to estimate the phase-space ratio of electrons in the semiconductor and the superconductor. %Since the disorder scattering rate in the semiconductor is a two-step process combining a rate $\gamma$ of exiting the well to the DSC \stutodo{must this be defined} followed by a re-entry with a rate proportional to $\gamma$, the disorder scattering rate scales as $\gamma^2$~\cite{lutchyn2012}.

The full disorder scattering rate $\tau_{d}^{-1}$ is the product of several factors that are written as
\begin{align}\label{taudwell}
\tau_{d}^{-1} &\sim  \gamma' \left( \frac{v_{F,sc}}{\lambda} \right) \left( \frac{\lambda}{\xi_{d}} \right) \left( \frac{1}{\Delta} \right) P' \left( \frac{m^{*}_{sm}\gamma}{k_{F,sc}^{2}}\right)  .
\end{align}
We explain the origin of the various factors in the rest of this paragraph. Here, $\gamma' \approx \mathrm{min}(\gamma,\Delta)$ is the collision rate with the barrier and the factor $v_{F,sc}/ \lambda =\tau_{mean}$ is the scattering rate in the superconductor. Only scattering events that occur close enough to the interface can cause transmission back into the superconductor, contributing a probability $\lambda/ \xi_{d}$. This process continues over the Andreev time $\tau_{sc,A}= 1/\Delta$. The factor $P'$ is the probability of crossing the barrier and the final factor $m^{*}_{sm}\gamma/k^{2}_{F,sc}$ is the phase space ratio from the previous paragraph, which is related to the likelihood that the particle will have a available state in the semiconductor. The collision rate $\gamma'$ is on the order of $\gamma$ in the weak coupling limit $\gamma\ll\Delta$. In the strong coupling limit $\gamma\gg\Delta$, based on arguments following Eq.~\ref{eq:tau_d}, the scattering rate is limited by the high probability Andreev reflection process so that $\gamma'\sim\Delta$.

Note that the transmission probability $P'$ on resonance is not the same as the transmission probability from inside the well. To understand this, consider replacing the superconductor by a clean normal metal with a Fermi velocity $v_{F,sc}$.
%If one calculates the quantum density of states inside the well, in the limit of a high barrier, one finds a total weight of 1 spread out over a range $\gamma$. The density of states in the well is thus $1/ \gamma$. 
Let us assume that the state in the well is tunnel broadened by $\gamma$, which also matches the decay rate of the electron in the well from 
leaking into the metal.
The flux of particles from the normal metal side, per unit energy, is proportional to $v_{F,sc} \times DOS \propto v_{F,sc}/v_{F,sc}=1$. The transition rate from the normal metal into the well over the energy range $\gamma$ would thus be $P'\gamma$. This matches the decay rate $\gamma$ of the electron which is trapped in the well if we assume $P'\sim 1$ .
Substituting this into Eq.~\ref{taudwell} and simplifying gives
\begin{align}\label{eq:taudlutchyn}
\tau_{d}^{-1} &\sim \frac{\gamma'\gamma m^*_{sm}}{\Delta m^*_{sc}\xi_{d} k_{F,sc}}.
\end{align}
This classical result is consistent (apart from numerical factors) to the result in Lutchyn et al~\cite{lutchyn2012} in the weak coupling limit where $\gamma'=\gamma$. However the strong coupling limit modifies this result to $\tau_{d}^{-1} \sim \frac{\gamma m^*_{sm}}{ m^*_{sc}\xi_{d} k_{F,sc}}$.

Thus, our results suggest that the suppression of bulk SC disorder demonstrated in the weak coupling limit in  Ref.~\cite{lutchyn2012} continues to apply for the quantum well case in strong coupling $\gamma\gtrsim\Delta$ when $k_{F,sc}\xi_{d}\gg 1$. Scattering in a quasi-one  dimensional nanowire case of width $W$ would add a phase-space factor of $W/k_{F,sm}$ from the estimate for a quantum well in the previous paragraph. An interesting case from the numerical stand-point, is the case of a purely two dimensional superconductor. In this case, we lose a factor of $k_{F,sm}/k_{F,sc}$ and we find that the disorder scattering is suppressed by $k_{F,sm}$ in the denominator of Eq.~\ref{eq:taudlutchyn} instead of $k_{F,sc}$. Thus, the suppression of disorder scattering at large $k_{F,sc}$ fails to apply in this strong coupling case which is consistent with numerical simulations in Ref.~\cite{cole2016}. This partly explains the discrepancy between the results of Ref.~\cite{lutchyn2012} and Ref.~\cite{cole2016} as being a result of different dimensionality of the superconductors rather than coupling. However,  the classical estimate fails to describe the numerics in Ref.~\cite{cole2016} for the strictly two-dimensional superconductor case likely because the $k_{F,sc}$ in the simulations were not large enough to enter the classical limit.

\section{Magnetic field enhancement of disorder scattering}%
\label{sec:magnetic-field}
The results in the previous sections show that the disorder scattering from a disordered superconductor with large Fermi momentum mismatch is dominated by normal reflection.
%The suppression of disorder scattering in the proximity-inducing superconductor originates from
This is because of the approximate conservation of quasiparticle momentum in the Andreev scattering process. We understand this effect classically as each hole retracing the time-reversed trajectory of the electron. Quantum mechanically, this is a result of the Cooperon contribution to the electron diffusion in the time-reversal symmetric case. The application of a small magnetic field breaks time-reversal symmetry and adds a momentum $\delta_e k=q_e\int dt \left(\dot{r}\times B\right)=q_e(\delta r\times B)$ over the electron part of the trajectory $r(t)$. Since both the charge of the electron $q_e$ as well as the displacement $\delta r$ changes sign, the change in momentum over the hole part of the trajectory is the same, i.e. $\delta_h k=\delta_e k$. Thus, the total change of momentum is $\delta k = 2q_e(\delta r\times B)$. The average magnitude of the momentum shift provides an estimate of the mean-free path from this magnetic-field induced scattering, i.e. $\lambda_{mag}=\expect{\delta k^{2}}^{-1/2}=1/(2 q_e B \sqrt{\expect{\delta r_{\perp}^2}})$, where $\delta r_{\perp}$ is the component of the displacement which is transverse to the magnetic field. This field can have a significant effect on topological superconductivity if the ratio
\begin{equation}
  \label{eq:magfield}
  \frac{\xi_0}{\lambda_{mag}} \sim 2 q_e B \xi_0 \sqrt{\expect{\delta r_{\perp}^2}}
\end{equation}
is comparable to unity, since $\xi_{0}/ \lambda_{mag} \sim (\Delta\tau_{mag})^{-1}$ where $\tau_{mag}$ is the magnetic scattering time.

Assuming, for a thin superconductor, that the mean-squared displacement is comparable to the thickness $d\sim \sqrt{\expect{\delta r_{\perp}^2}}$, the ratio $\xi_0/ \lambda_{mag}$ becomes significant (on the order of one) for a $d=10\si{nm}$ thick superconductor with $\xi_0 \simeq 500\si{nm}$ and with a relatively modest magnetic field of about $B\sim 0.2 \si{T}$. Thus, magnetic fields can eliminate the protection from disorder scattering in the superconductor that results from Fermi momentum mismatch. This enhanced scattering is consistent with numerical results described in the following section. However, the classical analysis suggests that the extra scattering momentum $\delta k$ is perpendicular to the applied magnetic field. This may provide protection in a magnetic field for single sub-band topological superconductors where the magnetic field is aligned with the topological superconductor. However, this requires more detailed quantum calculations that include a magnetic field.

 \section{Numerical evaluation of diagonal Andreev reflection}
 \label{sec:kwant}
 \newcommand{\CHISQ}{2.56}
 Topological superconductivity relies on the clean (or diagonal in momentum space) part of the Andreev reflection $|r_{eh}^{p}|^{2}$ defined in Eq.~\ref{eq:beenakker_approx_prob}. The random matrix  results of Secs.~\ref{sec:full-calculation} and Appendix \ref{sec:offdiagonal} show that this diagonal part of the Andreev reflection dominates the scattering process in the time-reversal invariant case with large momentum mismatch. The argument in the last section shows that the diagonal Andreev reflection is suppressed in the presence of an orbital magnetic field. These results were based on a combination of random matrix theory and semiclassical dynamics. In this section, we perform a direct numerical simulation of Andreev reflection using Kwant \cite{kwant}. We use a 2D square bilayer lattice of width $200\si{nm}$ to simulate the effective model shown in Fig.~\ref{fig:eff-model} (the bilayer allows for an in-plane magnetic field, see next paragraph). The lattice constant is $a=1\si{nm}$ with the distance between bilayers being $a_{z}=20\si{nm}$. We take the effective mass to be $1.4m_{e}$ in the superconductor and $0.02m_{e}$ in the semiconductor, using a chemical potential of $\mu_{sc} = 98.6\si{meV}$ in the superconductor and $\mu_{sm}=20.48\si{meV}$ in the superconductor. We introduce a pairing potential $\Delta=1\si{mev}$ into the superconductor. This large gap is not physical but reduces the coherence length to $\xi_{0}=2 a \sqrt{\mu_{sc} t}/ \pi \Delta \approx 33.0\si{nm}$, allowing numerics with smaller systems. To implement the adiabatic SM-N interface, we smoothly change the potential over a $L_{sm}=2000$ transition region which is attached to a $L_{sc}=4 \xi_{0}$ superconductor with Gaussian, uncorrelated disorder of strength $V$, giving a mean free path of $\lambda =a \sqrt{\mu_{sc}t^{3}}/V^{2}$. In the $V=0$ case, the adiabatic barrier has a reflection probability on the order of $10^{-4}$. Clean leads are attached to the left and right sides of the system with $N_{sm}=8$ propagating modes in the semiconducting lead and $N=164$ propagating modes in the superconducting lead. We simulate $10$ realizations of $50$ disorder strengths which are shown in Fig.~\ref{fig:kwant}. The theoretical calculation gives a reduced chi-squared statistic of $\chi^{2}/\mathrm{DOF} = \CHISQ$.

As discussed in Sec.~\ref{sec:magnetic-field}, the orbital effects of a transverse magnetic field $B$ enhance the disorder scattering from the superconductor in the classical approximation. In the tight-binding system, we model this effect using a Peierls substitution, replacing the transverse hopping $t_{x}$ with $t_{x}\exp(iq_{e}A_{x}a a_{z} \tau_{z})$ where $A_{x}$ is the longitudinal component of the vector potential and $\tau_{z}$ is a Pauli matrix in Nambu space. To define a transverse in-plane field, $A_{x}$ must vary in the $z$ direction, which motivates the use of a bilayer system. Therefore, we substitute $t_{x}\mapsto t_{x}\exp(i q_{e} B a a_{z} \tau_{z})$ in one of the layers of the model. This field has a strength of $B\approx3.15\Phi_{0}/ \xi_{0} a_{z}$, producing a cyclotron radius $r_{c}=m^{*}_{sc}v_{F,sc}q_{e}B\approx 127.8\si{nm}$. The physical value of $B$ is quite high, corresponding to $B=10\si{T}$, since the coherence length is nonphysically small. However, the inverse relationship between the coherence length and the field strength shown in Eq.~\ref{eq:magfield} indicates that a system with $\xi_{0}\simeq 500$ would experience similar disorder enhancement at $B\sim0.7\si{T}$.

 \begin{figure}[h]
  \centering
  \includegraphics[width=0.9\columnwidth]{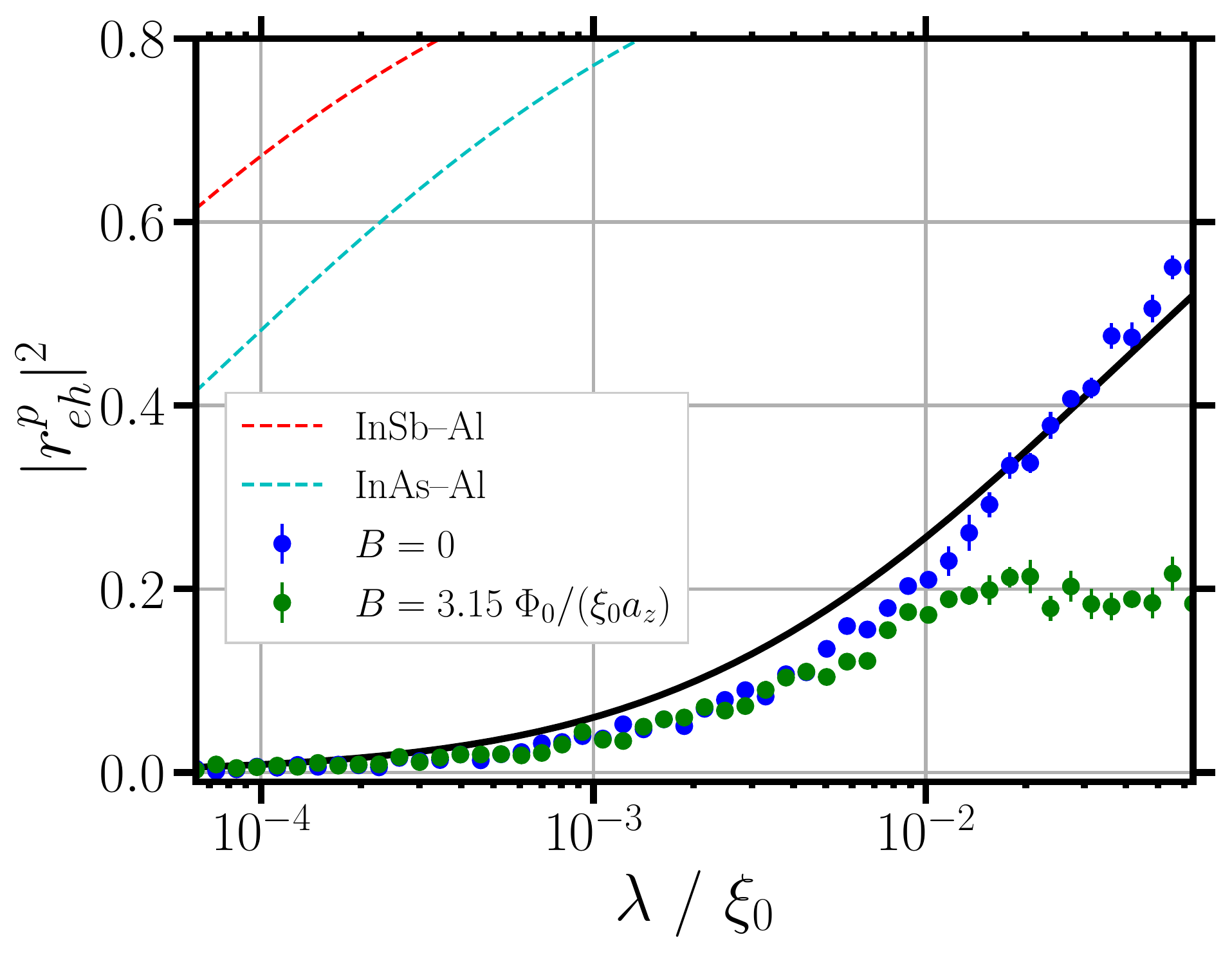}
  \caption{\label{fig:kwant}Numerical simulation of perfect Andreev reflection probability $|r_{eh}^{p}|^{2}$ as a function of normal region conductance. Results are averaged over $10$ realizations of Gaussian on-site disorder with model parameters $m_{sc}^{*}=1.4m_{e}$, $m_{sm}^{*}=0.02m_{e}$, $\mu_{sc} = 98.6\si{meV}$, $\mu_{sm}=20.48\si{meV}$, $\Delta=1\si{mev}$, $a_{z}=20\si{nm}$. The number of propagating modes is $N_{sm}=8$ in semiconductor and $N=164$ in the superconductor. This simulation is compared to theoretical result in Eqs.~\ref{eq:beenakker_approx_prob} and \ref{eq:diffusive} with and without a transverse in-plane magnetic field in units of the flux quantum, the clean superconducting coherence length and the bilayer separation. The theoretical result gives $\chi^{2}/\mathrm{DOF} = \CHISQ$ in the $B=0$ case. These results are compared to theoretical predictions using the physical values of the InAs-Al and InSb-Al junctions (see Sec.~\ref{sec:physical-realization}).}
\end{figure}

The numerical results for the no-field case shown in Fig.~\ref{fig:kwant} agrees quite well with the random matrix result in Eqs.~\ref{eq:diffusive} and \ref{eq:beenakker_approx_prob}, which is shown by the solid black line. Realistic materials, such as the InAs-Al and InSb-Al junctions discussed in Sec.~\ref{sec:physical-realization},
have a higher degree of Fermi momentum mismatch and exhibit much more disorder suppression than numerically feasible systems. Application of a magnetic field (green dots in Fig.~\ref{fig:kwant}), even with a cyclotron radius larger than the bilayer separation, is enough to suppress diagonal Andreev reflection substantially.

\section{Conclusion}
\label{sec:discussion}
Topological superconductors are rather sensitive to disorder scattering and require relatively clean systems where the mean-free path exceeds the superconducting coherence length~\cite{motrunich2001,brouwer2011topological,sau2012,ahn2021estimating}. While semiconductor fabrication has been improved to the point of being able to achieve mean-free paths over a micron, the mean free path of the superconductor is typically much shorter because the superconductor is not epitaxially grown. Even in the case of nanowires where the superconductor is epitaxial~\cite{krogstrup2015}, the mean-free path in the superconductor is limited by surface scattering.  This poses a potential challenge in systems where strong coupling to the superconductor is often needed for creating robust topological gaps.

Using a combination of scattering theory and random matrix calculations, we find in Sec.~\ref{sec:full-calculation} that the effects of disorder scattering in the bulk superconductor are suppressed as the ratio of the Fermi wave-length in the semiconductor to that in the superconductor. This regime is also characterized by near perfect Andreev reflection, which can be used as a signature of being in this low disorder scattering regime. We calculate the maximum disorder (minimum mean free path) in realistic materials and find a higher disorder tolerance than previously expected~\cite{lutchyn2012,hui2015,cole2016}: In InAs-Al, the mean free path in the superconductor must exceed lengths of $\sim 1 \si{nm}$ to generate a topological gap. Our results might be a possible explanation for why many semiconductor/superconductor experiments~\cite{hart2019, ren2019} appear to show high Andreev reflection probabilities relative to metal/superconductor interfaces, where this effect rarely occurs~\cite{lee2019}.

Interestingly, the results of our scattering matrix calculation turn out to be consistent with the classical estimate (Sec.~\ref{sec:classical-estimate}) for the scattering rate at the leading order in the ratio of the Fermi momenta. The classical estimate provides an understanding of disorder scattering suppression as a result of a reduction in phase space, making disorder scattering highly unlikely without affecting Andreev scattering. We find that the same classical approach can show that previous results on superconducting scattering~\cite{lutchyn2012,potter2011} continues to be valid in the moderately strong superconducting coupling limit as long as the Fermi momenta are sufficiently different. The apparent discrepancy of these works with the numerical result~\cite{cole2016} where strong disorder scattering is noted can be attributed to the purely two dimensional nature of the superconducting model.

Finally, we show in Sec.~\ref{sec:magnetic-field} that the introduction of magnetic fields can significantly enhance disorder effects by introducing scattering into the Andreev reflection, which is otherwise momentum conserving.  However, the magnetic field must lie perpendicular to the momentum scattering. A similar effect maybe expected from the combination of spin-orbit scattering and in-plane Zeeman field, which has not been explicitly studied here. The effect of both these contributions to disorder requires a careful treatment of time-reversal and spin-conservation breaking, which is left for future studies. While future work is needed to understand the proximity effect from spin-orbit coupled superconductors in a magnetic field, our work shows that as a matter of principle, even the normal state mobility constraints on superconductors that are used to realize topological superconducting phases by proximity effect are not strongly restrictive if the semiconductor has low density. Demonstration of near perfect Andreev reflection at a semiconductor-superconductor interface is a good indicator that such a condition has been met in the absence of a magnetic field. \\

\noindent \textit{Note added--} After the completion of this manuscript, two recent preprints appeared with some connections to the current work.  One preprint investigated numerically, through realistic non-perturbative first principles simulations, the effects of disorder in the superconductor on the proximity effect in the hybrid superconducting thin film and semiconducting nanowire platform in the presence of Fermi surface mismatch, concluding, in agreement with us, that the disorder in the superconducting film mostly has negligible detrimental effects on the proximity effect \cite{stanescu2022}.  The second preprint presents extensive experimental results on Al-InAs superconductor-nanowire platforms, providing direct evidence for a topological gap in InAs in spite of considerable disorder (mean free path ~10 nm) in the Al film, again in agreement with our theoretical conclusion in the current work \cite{aghaee2022}.

\begin{acknowledgements}%
We thank Anton Akhmerov for valuable discussions that alerted us to the importance of Fermi momentum mismatch in suppressing disorder scattering.
We also thank David Goldhaber Gordon for discussions of controlling disorder scattering at the semiconductor/superconductor interface that motivated this work. Stuart Thomas thanks the Joint Quantum Institute at the University of Maryland for support through a JQI fellowship. JS acknowledges NSF DMR – 1555135 for support.
This work is also supported by the Laboratory for Physical Sciences through its continuous support of the Condensed Matter Theory Center at the University of Maryland.
\end{acknowledgements}%

\appendix

\section{Calculation Details}
\label{sec:calculation-details}
Using the Haar measure, we can calculate the integrals of unitary matrix moments as Weingarten functions~\cite{collins2006,brouwer1996}:
\begin{align}
  \label{eq:wg_def}
  &\left\langle U_{i_{1}j_{1}} \cdots U_{i_{n}j_{n}}U^{*}_{i'_{1}j'_{1}}\cdots U^{*}_{i'_{n}j'_{n}}\right\rangle \nonumber \\
  &\qquad =\sum_{P,P'} \delta_{i_{1}i'_{P(1)}}\cdots \delta_{i_{n}i'_{P(n)}}\delta_{j_{1}j'_{P'(1)}}\cdots\delta_{j_{n}j'_{P'(n)}} \nonumber \\
    &\qquad\qquad \times \Wg\left(P^{-1}{P'},\, N\right)
\end{align}
where $P$ and $P'$ sum over all permutations in the symmetric group $\mathrm{Sym}(n)$. Weingarten functions decrease asymptotically with respect to the matrix dimension $N$ and thus higher order terms in $N$ are negligible for multichannel metals \cite{weingarten1978}. For leading order in $N$, the Weingarten function of a permutation $\sigma$ can be expressed as a product over cycle lengths $C_{i}$, including one-cycles:
\begin{equation}
  \label{eq:wg}
  \Wg(\sigma,N) = \prod_{i}\frac{(-1)^{ C_{i} -1 } c_{ C_{i}-1}}{N^{2C_{i} -1}} + O(N^{-n-1})
\end{equation}
where $c_{n}\equiv(2n)!/n!(n+1)!$ is the $n$-th Catalan number.

%  By expanding the indices and grouping the $U$ factors together, we isolate the unitary moments:
% \begin{equation}
%   \label{eq:expansion}
%   \Gamma_{n}\sim \sum_{\{i\},\{j\}} \left\langle\prod^{2n+2}_{\substack{\mathrm{even} \\ \mu=2}}U_{i_{\mu}i_{\mu-1}}U_{i_{\mu}i_{\mu+1}}U^{*}_{j_{\mu}j_{\mu-1}}U^{*}_{j_{\mu}j_{\mu+1}}\right\rangle.
% \end{equation}
%
%

Integrating these moments yields a product of the Weingarten function and Kronecker deltas:
\begin{align}
  \label{eq:gamma-expanded}
  \Gamma_{n}= \sum_{\{i\},\{j\}}&t_{i_{1}} \delta_{i_{2n+3}j_{1}}t_{j_{1}} \delta_{j_{2n+3},i_{1}} \nonumber \\
                           \times&\prod^{2n+2}_{\substack{\mathrm{even}\\ \mu=2}} \sqrt{R_{i_{\mu}}  R_{j_{\mu}}}\prod^{2n+1}_{\substack{\mathrm{odd}\\ \mu=3}} r_{i_{\mu}} r_{j_{\mu}} \nonumber \\
  \times&\sum_{P,P'}\mathrm{Wg}(P^{-1}P',N)\prod^{2n+2}_{k=1}\delta_{a_{k}\alpha_{P(k)}}\delta_{b_{k}\beta_{P'(k)}}
\end{align}
where $r_{i}$, $t_{i}$ and $R_{i}$ are the diagonal elements of $r_{sm}$, $t_{sm}$ and $R_{i}$ respectively; $P$ and $P'$ are elements of $\mathrm{Sym}(2n+2)$; and $a$ and $b$ ($\alpha$ and $\beta$) are the left and right indices of $U$ ($U^{*}$) respectively, defined explicitly as $a_{k}=i_{2\lfloor (k+1)/2 \rfloor}$, $b_{k}=i_{2\lfloor k/2\rfloor+1}$, $\alpha_{k}=j_{2\lfloor (k+1)/2 \rfloor}$ and $\beta_{k}=j_{2\lfloor k/2\rfloor+1}$.

To simplify Eq.~\ref{eq:gamma-expanded}, we make a series of observations. First, any Kronecker delta factor which shares indices with both a $t_{i}$ factor and a $r_{i}$ factor sums to zero since the nonzero entries of these diagonals are disjoint. Furthermore, each delta term preserves the parity of indices, mapping even to even and odd to odd. Therefore we can assume every non-zero term in the sum over $P,P'$ contains either $\delta_{i_{1}j_{1}}$ and $\delta_{i_{2n+3}j_{2n+3}}$ or $\delta_{i_{1}j_{2n+3}}$ and $\delta_{i_{2n+3}j_{1}}$. In other words, either $P'(1)=1$ and $P'(2n+3)=2n+3$ or $P'(1)=2n+3$ and $P'(2n+3)=1$.

Since we are chiefly concerned with multichannel metals, we consider only the leading term in $N$. With this in mind, we note that each index (other than the first and last) comes in a pair. Therefore each index has two Kronecker delta factors relating it to two other indices. We can view this structure as a graph of disjoint cycles. When summing over all $i$ and $j$ terms, the Kronecker deltas contract the indices, resulting in a number of summations over $N$ equal to the number of Kronecker delta cycles. Since each summation (other than the first and last indices) contributes a factor proportional to $N$, the dominant permutations are those that create the most cycles. Specifically, these configurations manifest as those that preserve the pairs of indices such that each cycle consists of only two Kronecker delta terms.

Finally, we note that the Weingarten function $\Wg (P^{-1}P',N)$ has the highest power in $N$ when the permutation $P^{-1}P'$ is the product of 1-cycles--- i.e. the identity permutation. Therefore we assume $P=P'$ for any relevant terms.

Using these classifications of the dominant terms, we consider the previously mentioned case where $P'(1)=1$ and $P'(2n+3)=2n+3$. Since $P=P'$, we find that $P(1)=1$, leading to a $\delta_{i_{2}j_{2}}$ term. Since each pair must map to itself in the leading $N$ term, the next index of $i_{2}$ must map to the next index of $j_{2}$, such that $P(2)=2$. Therefore $P'(2)=2$ , mapping $i_{3}$ to $j_{3}$. We can continue this reasoning to inductively show that $P=P'$ is the identity. In the $P'(1)=2n+3$ and $P'(2n+3)=1$ case, similar reasoning shows that $P=P'$ is equal to $[(1,2n+3)(2,2n+2)\cdots(n+1,n+3)]$, or the mapping $i_{k} \mapsto j_{2n+4-k}$. We refer to this case as the exchange permutation.

Pulling out the sum over permutations, we separate $\Gamma_{n}$ into terms $\Gamma^{\mathrm{id}}_{n}$ and $\Gamma^{\mathrm{ex}}_{n}$ representing the identity permutation and exchange permutation respectively. Contracting the delta functions in the identity term gives
\begin{align*}
 \Gamma^{\mathrm{id}}_{n} &=\Wg\left(\mathbb{1},N\right) \sum_{\{i\}}t_{i_{1}}t_{i_{2n+3}} \delta_{i_{1}i_{2n+3}} \prod^{2n+2}_{\substack{\mathrm{even}\\ \mu=2}}R_{i_{\mu}} \prod^{2n+1}_{\substack{\mathrm{odd}\\ \mu=3}}r_{i_{\mu}} \\
 &= \frac{1}{N^{2n+2}} \left(N_{sm}\right)\left(N \overline{R_{N}}\right)^{n+1} \left(N - N_{sm}\right)^n \\
  &= \frac{N_{sm}}{N}\, {\overline{R_{N}}}^{n+1} \left(1-\frac{N_{sm}}{N}\right)^{n}.
\end{align*}
where $\overline{R_{N}}$ is the average conductance of a single channel in the normal metal. The exchange term $\Gamma^{ex}_{n}$ evaluates to the same expression with an extra factor of $N_{sm}$ coming from the additional sum over $t_{i_{2n+3}}$. Assuming $N_{sm}\gg 1$, the identity term becomes negligible and $\Gamma_{n}$ reduces to Eq.~\ref{eq:gamma-final}.

Note that similar arguments apply in the diagrammatic method \cite{brouwer1996}, yielding the same result.

\section{Generalization to non-square transmission matrices}
\label{sec:beenakker_generalization}
From Ref.~\cite{beenakker1992}, the Andreev reflection matrix is
\begin{equation}
  \label{eq:reh_expansion}
  r_{eh}=-i t_{LR}\left(1+r_{R}^{\dag}r_{R}\right)^{-1}t_{LR}^{\dag}
\end{equation}
where $t_{LR}$ and $r_{R}$ describe the right-to-left transmission and right-side reflection of the SM-N interface respectively. Noting that $t_{LR}$ is rectangular with dimension $N_{sm}\times N$, we use the singular value decomposition (SVD) to write it as $t_{LR}=W^{T}\mathcal{T}V$ where $W$ and $V$ are unitary matrices and $\mathcal{T}$ is a real rectangular diagonal matrix. By the unitarity of the full scattering matrix, $t_{LR}^{\dag}t_{LR}=1-r_{R}^{\dag}r_{R}$ and $t_{LR}^{*}t_{LR}^{T}=1-r_{L}^{\dag}r_{L}$. The SVD then implies $\mathcal{T}^{T}\mathcal{T}=1-R_{R}$ and $\mathcal{T}\mathcal{T}^{T}=1-R_{L}$ where $R_{R}$ and $R_{L}$ are the diagonal matrices of right and left reflection eigenvalues, respectively. Since $\mathcal{T}$ is rectangular diagonal, $R_{R}=\mathrm{diag}(R_{L},\mathbb{0})$, assuming that $\mathrm{dim}(R_{R})>\mathrm{dim}(R_{L})$. Furthermore, the unitarity relation connects the SVD matrices $W$ and $V$ with the polar decomposition such that $r_{L}=W^{T}\sqrt{R_{L}}W$ and $r_{R}=V^{T}\sqrt{R_{R}}V$, allowing us to rewrite $r_{eh}$ as
\begin{align}
  % r_{eh}=&-i W^{\dagger}\mathcal{T}V \left(1+V^\dagger R_{R} V\right)^{-1}V^{\dagger} \nonumber \\
  % &\qquad \times \mathcal{T}^{T}\mathcal{T}V\left(1+V^{\dagger}R_{R}V\right)^{-1}V^{\dagger}\mathcal{T}^{T}W \nonumber \\
    r_{eh}=&-iW^{\dagger}\mathcal{T} \left(1+R_{R}\right)^{-1}\mathcal{T}^{T}W\nonumber \\
    =&-iW^{\dagger}\left(\frac{1-R_{L}}{1+R_{L}}\right) W \nonumber \\
    =&-i\left(\frac{1-r_{L}r_{L}^{\dagger}}{1+r_{L}r_{L}^{\dagger}}\right).
\end{align}
This gives the derivation of Eq.~\ref{eq:reh}.

\section{Off-diagonal reflection}
\label{sec:offdiagonal}
%The addition of a superconductor to the right hand side of this system causes Andreev reflection to occur, described by the Andreev reflection matrix $r_{eh}$. We assume that the pure N-SC interface exhibits perfect Andreev reflection at zero energy, or $r^{A}_{eh}=-i \times \mathbb{1}$. Therefore the matrix composition
Following Beenakker~\cite{beenakker1992}, the Andreev reflection matrix $r_{eh}$ for zero-energy Andreev reflection of electrons coming from the left in Fig.~\ref{fig:eff-model} is given by:
\begin{align}
  \label{eq:beenakker_sm}
 r_{eh} = -i t_{LR} \left( 1+r_{R}^{*}r_{R}\right)^{-1} t_{RL}^*.
\end{align}
which describes the total Andreev reflection of the SM-N-SC system.
Here, $r_{R}$ is the right-side reflection and $t_{LR(RL)}$ is the right-to-left (left-to-right) transmission of the SM-N junction. In this case, we assume $t_{LR(RL)}$ is a $N_{sm}\times N$ ($N\times N_{sm})$ rectangular matrix, only mapping to allowed modes. Since the scattering matrix is symmetric, $t_{LR}=t_{RL}^T$ and $r^*_R=r_R^\dag$. Using the singular value decomposition and the unitarity of the scattering matrix, we find Eq.~\ref{eq:reh}.

To calculate the magnitude of off-diagonal reflection, we first find the variance of the reflection eigenvalues ${R_{L}}_{i}$. Using similar reasoning to the calculation in Sec.~\ref{sec:full-calculation}, we calculate the second moment of the reflection eigenvalues using the trace $\overline{R_{L}^{2}}=(1/N_{sm})\left\langle\tr r_L r_L^\dag r_L r_L^\dag\right\rangle$. In this case, we cannot reduce the calculation to a single sum. Each of the four $r_{L}$ matrices contribute an infinite sum with one sum reduced to enforce equal numbers of $U$ and $U^*$ factors, resulting in a three-dimensional infinite series. Following the logic of the first moment derivation, we break the expression into terms $\Delta_{a,b,c,d}$:
\begin{equation}
  \label{eq:2}
  \left\langle \tr r_L r_L^\dag r_L r_L^\dag  \right\rangle = \sum_{a,b,c}^{\infty} \Delta_{a,b,c,a+c-b}.
\end{equation}
In the second moment case, the terms take the form
\begin{widetext}
\begin{align*}
  \Delta_{a,b,c,d}=&
    \sum_{\{i\},\{j\},\{k\},\{l\}}^{N}  t_{i_{1}}\delta_{i_{2a+3}j_{1}}t_{j_{1}} \delta_{j_{2b+3}k_{1}}t_{k_{1}} \delta_{k_{2c+3}l_{1}}t_{l_{1}} \delta_{l_{2d+3}i_{1}}\\
    &\qquad \times \left(
    \prod^{2a+2}_{\substack{\mathrm{even}\\\mu=2}} \sqrt{R_{i_{\mu}}}
    \prod^{2b+2}_{\substack{\mathrm{even}\\\mu=2}} \sqrt{R_{j_{\mu}}}
    \prod^{2c+2}_{\substack{\mathrm{even}\\\mu=2}} \sqrt{R_{k_{\mu}}}
    \prod^{2d+2}_{\substack{\mathrm{even}\\\mu=2}} \sqrt{R_{l_{\mu}}}
    \right)\left(
    \prod^{2a+1}_{\substack{\mathrm{odd}\\\mu=3}} r_{i_{\mu}}
    \prod^{2b+1}_{\substack{\mathrm{odd}\\\mu=3}} r_{j_{\mu}}
    \prod^{2c+1}_{\substack{\mathrm{odd}\\\mu=3}} r_{k_{\mu}}
    \prod^{2d+1}_{\substack{\mathrm{odd}\\\mu=3}} r_{l_{\mu}}\right) \\
    &\qquad \times  \left\langle
    \prod^{2a+2}_{\substack{\mathrm{even}\\\mu=2}} U_{i_{\mu}i_{\mu-1}} U_{i_{\mu}i_{\mu+1}}
    \prod^{2b+2}_{\substack{\mathrm{even}\\\mu=2}} U^{*}_{j_{\mu}j_{\mu-1}} U^{*}_{j_{\mu}j_{\mu+1}}
    \prod^{2c+2}_{\substack{\mathrm{even}\\\mu=2}} U_{k_{\mu}k_{\mu-1}} U_{k_{\mu}k_{\mu+1}}
    \prod^{2d+2}_{\substack{\mathrm{even}\\\mu=2}} U^{*}_{l_{\mu}l_{\mu-1}} U^{*}_{l_{\mu}l_{\mu+1}}
    \right\rangle.
\end{align*}
\end{widetext}

Recall that nonzero terms in Weingarten sum must map $t_{i}$ factors to other $t_{i}$ factors. In the first moment calculation, we inductively demonstrated that $P(1)=1$ implies $P'(2a+3)=2a+3$. Similarly in the exchange permutation, $P'(1)=2a+3$ implies $P'(2a+3)=1$. In order for these iterations to map $t_{i}$ factors together, $a$ must equal $b$. Generally, in the second moment case, we also permit permutations which map from $i$ indices to $l$ indices and $k$ indices to $j$ indices. In this scenario, similar reasoning enforces $b=c$. Therefore all nonzero terms obey $a=b$ or $b=c$. Since $\Delta_{a,b,c,d}$ is symmetric under $a\leftrightarrow c$ transformation \footnote{This substitution leads to equivalent changes in $P$ and $P'$ which cancel in the expression $P^{-1}P'$.} we write
\begin{equation}
  \label{eq:2m_fullsum}
  \left\langle \tr r_L r_L^\dag r_L r_L^\dag  \right\rangle  = 2 \sum_{m\neq n}^{\infty} \Delta_{m,n} + \sum_{n}^{\infty} \Delta_{n,n},
  % \tr[r_L r_L^\dag r_L r_L^\dag]  &= 2 \sum_{m, n}^{\infty} \Delta_{m,m,n,n} - \sum_{n}^{\infty} \Delta_{n,n,n,n}
\end{equation}
where $\Delta_{m,n}\equiv \Delta_{m,m,n,n}$, separating the infinite sum into asymmetric and symmetric terms.

% \begin{widetext}
% \begin{align*}
%    \Delta_{m,m,n,n}=& \sum_{\{i\},\{j\},\{k\},\{l\}}^{N}  t_{i_{1}}\delta_{i_{2m+3}j_{1}}t_{j_{1}} \delta_{j_{2m+3}k_{1}}t_{k_{1}} \delta_{k_{2n+3}l_{1}}t_{l_{1}} \delta_{l_{2n+3}i_{1}}\\
%   &\qquad \times \left(\prod^{2m+2}_{\mathrm{even}\;\mu=2}\sqrt{R_{i_{\mu}}R_{j_{\mu}}} \prod^{2n+2}_{\mathrm{even}\;\mu=2}\sqrt{R_{k_{\mu}}R_{l_{\mu}}}\right)  \left(\prod^{2m+1}_{\mathrm{odd}\;\mu=3}r_{i_{\mu}}r_{j_{\mu}} \prod^{2n+1}_{\mathrm{odd}\;\mu=3}r_{k_{\mu}}r_{l_{\mu}}\right) \\
% &\qquad \times  \left\langle\prod^{2m+2}_{\mathrm{even}\;\mu=2} U_{i_{\mu}i_{\mu-1}} U_{i_{\mu}i_{\mu+1}} U^{*}_{j_{\mu}j_{\mu-1}} U^{*}_{j_{\mu}j_{\mu+1}} \prod^{2n+2}_{\mathrm{even}\;\mu=2}U_{k_{\mu}k_{\mu-1}} U_{k_{\mu}k_{\mu+1}} U^{*}_{l_{\mu}l_{\mu-1}} U^{*}_{l_{\mu}l_{\mu+1}}\right\rangle.
% \end{align*}
% \end{widetext}
In the asymmetric term there are four ways to link the $t_{i}$ factors---identity and exchange permutations between $i \leftrightarrow j$ and $k \leftrightarrow l$---forming the direct product group $\mathrm{Sym}(2)\times \mathrm{Sym}(2)$. We associate each element of this group with $N_{sm}$ raised to the number of disjoint cycles, indicating the contribution of the $t_{i}$ terms. Summing these terms gives $N_{sm}^{3}+N_{sm}^{2}+N_{sm}^{2}+N_{sm}\approx N_{sm}^3$ for $N_{sm}\gg1$. Following the Sec.~\ref{sec:full-calculation} calculation, we find
\begin{equation}
\label{eq:delta}
    \Delta_{m,n} =   \frac{N_{sm}^3}{N^{2}} \; {\overline{R_{N}}}^{m+n+2} \left(1-\frac{N_{sm}}{N}\right)^{m+n}.
\end{equation}

The symmetric term $\Delta_{n,n}$ introduces a third degree of freedom for the permutation $P'$, incorporating those that map $i$ indices to $l$ indices and $j$ indices to $k$ indices ($P'\in\mathrm{Sym}(2)\times\mathrm{Sym}(2)\times\mathrm{Sym}(2)$). This contributes a factor of $2$ compared to the asymmetric terms so that $\Delta_{m=n}=2\Delta_{m\neq n}$. Therefore the full sum becomes $\overline{R_{L}^2}=2\sum_{m,n}\Delta_{m,n}$ with $\Delta_{m,n}$ defined in Eq.~\ref{eq:delta},
evaluating to
\begin{equation*}
  \overline{R_{L}^{2}} = 2\left(\frac{Z^{2}}{1+Z^{2}}\right)^{2},
\end{equation*}
using the definition of $Z^{2}$ from Eq.~\ref{eq:rmat}.

Relating our results to the SM-N-SC junction, we compare the magnitude of off-diagonal Andreev reflection to diagonal reflection using Eq.~\ref{eq:beenakker}. Seeing that the second moment of the reflection eigenvalues ${R_{L}}_{i}$ approaches $0$ in the large $N$ limit, we can approximate $r_{eh}$ as
\begin{equation}
  \label{eq:beenakker_approx}
  r'_{eh} = -i\left( \frac{1-\overline{R_{L}}}{1+\overline{R_{L}}} \right) \times \mathbb{1}
\end{equation}
by assuming that ${R_{L}}_{i}$ are uniform (i.e. ${R_{L}}_{i}=\overline{R_{L}}$ ). The accuracy of this approximation is quantified by the trace $\tr r_{od}r_{od}^{\dag}$ where $r_{od}\equiv r_{eh} - r'_{eh}$ is the off-diagonal reflection matrix. By Taylor expanding around small $\overline{R_{L}}$ and $r_{L}r_{L}^{\dagger}$ we have
\begin{align*}
\tr r_{od}r_{od}^{\dag} &\approx \tr\left( 2 \overline{R_{L}} - 2 r_{L}r_{L}^{\dag} \right)^{2} \\
% &= 4 \left( N_{sm}\overline{R_{L}}^{2} + \tr\left(r_{L}r_{L}^{\dag}\right)^{2} - 2 \overline{R_{L}} \tr r_{L}r_{L}^{\dag} \right) \\
& = 4 N_{sm} \left( \overline{R_{L}^{2}} - \overline{R_{L}}^{2}\right)
\end{align*}
which is proportional to the variance of the eigenvalues ${R_{L}}_{i}$. Using the results from the first and second moment calculations, we find the probability of off-diagonal $e\rightarrow h$ reflection to be
\begin{equation*}
\left|r_{od}\right|^2 = 4 \left( \frac{Z^{2}}{1+Z^{2}} \right)^{2},
\end{equation*}
demonstrating that as $\overline{R_{L}}$ goes to $0$ (i.e. $T\rightarrow 1$), the Andreev reflection becomes diagonal. We therefore find that the off-diagonal reflection is much smaller than the normal reflection and that $r_{eh}$ is well approximated by the scalar matrix $r'_{eh}$, resulting in Eq.~\ref{eq:beenakker_approx_prob}. This expression gives a conductance approximation
\begin{equation*}
    G_{NS} \approx\frac{4 e^{2}}{h}\, \frac{N_{sm}}{ \left(1+2 Z^{2}\right)^{2}},
\end{equation*}
refining the value in Eq.~\ref{eq:conductance_bound_Z2}.

\section{Thin Shell Calculation}
\label{sec:thinshelldetails}
In Sec.~\ref{sec:thinshell}, the average transmission eigenvalue of the combined metal-barrier region is
\begin{align}
  \overline{T_{NB}} &= \Gamma \tr \left( 1-V^{\dag} \sqrt{\left( 1-T_{N}' \right) \left( 1-\Gamma \right)}V^{*}\right)^{-1}\nonumber \\
  &\times V^{\dag} T_{N}'V \left( 1-V^{T} \sqrt{\left( 1-T_{N}' \right) \left( 1-\Gamma \right)}V\right)^{-1}.
\end{align}
where $T_N$ is the diagonal matrix of transmission eigenvalues and $V$ is a Haar-distributed unitary matrix. Expanding out the geometric series gives $\overline{T_{TS}}=\Gamma \sum_{n=0}^\infty \Delta_n$ where
% \begin{align}
%   \Delta_{n} =& \left(V^{\dag} \sqrt{\left( 1-T_{N}' \right) \left( 1-\Gamma \right)}V^{*}\right)^{n}V^{\dag}\nonumber \\
%   &\times T_{N}'V \left(V^{T} \sqrt{\left( 1-T_{N}' \right) \left( 1-\Gamma \right)}V\right)^{n}.
% \end{align}
\begin{align}
  \label{eq:thinshell-expanded}
  \Delta_{n}=& \left(1-\Gamma\right)^{n}\sum_{\{i\},\{j\}} \delta_{i_{2n+2}j_{1}} T_{j_{1}} \delta_{j_{2n+2},i_{1}} \nonumber \\
                           \times&\prod^{2n}_{\substack{\mathrm{even}\\ \mu=2}} \sqrt{\left(1-T_{i_{\mu}}\right)\left(1-T_{j_{\mu}}\right)}\nonumber \\
  \times&\sum_{P,P'}\mathrm{Wg}(P^{-1}P',N)\prod^{2n+1}_{k=1}\delta_{a_{k}\alpha_{P(k)}}\delta_{b_{k}\beta_{P'(k)}}
\end{align}
after integrating over $V$. As in Appendix \ref{sec:calculation-details}, $a_{k}=i_{2\lfloor (k+1)/2 \rfloor}$, $b_{k}=i_{2\lfloor k/2\rfloor+1}$, $\alpha_{k}=j_{2\lfloor (k+1)/2 \rfloor}$ and $\beta_{k}=j_{2\lfloor k/2\rfloor+1}$.

Also following Appendix \ref{sec:calculation-details}, we recall that the Weingarten function is maximum in $N$ when $P=P'$; however unlike in Sec.~\ref{sec:calculation-details}, every cycle in Eq.~\ref{eq:thinshell-expanded} contributes a factor of $N$. Therefore the leading term is given by the permutation which creates the most cycles, namely the identity permutation ($P=P'=\mathbb{1}$). Summing this term over the $i$ and $j$ indices yields $\Delta_{n}= T_{N} \left( 1-T_{N} \right)^{n} \left( 1-\Gamma \right)^{n}$. The full geometric sum then becomes $\overline{T_{NB}}=(\overline{T_{N}'}^{-1}+\Gamma^{-1}-1)^{-1}$. A similar result is found in Ref.~\cite{brouwer1996}.

% \section{Generalization of $G_{NS}$ to rectangular matrices}
% In Sec.~\ref{sec:full-calculation} we follow \cite{beenakker1992} to calculate $G_{NS}$ in terms of the reflection eigenvalues ${R_{L}}_{i}$, however the original derivation assumes equal numbers of modes in the left and right leads. The key step from this assumption is

% \begin{equation*}
%   t_{LR}^{\dag}t_{LR} = 1-r_{R}^{\dag}r_{R}
% \end{equation*}
% \begin{equation*}
%   S = \mat{r_{L} & t_{LR} \\ t_{LR}^{T} & r_{R}}
% \end{equation*}
% Since $S$ is unitary, $r_{L}$
% Using the singular value decomposition

\bibliography{bib.bib}

\end{document}